\documentclass[usenatbib]{mn2e}
\usepackage{textcomp,graphicx,amsmath,pdflscape,amssymb,setspace}
\usepackage{amsmath,amssymb}
\usepackage{mathptmx}
\usepackage{float}
         
\newcommand\apj{{ApJ}}%
\newcommand\ao{{Appl.~Opt.}}%
\newcommand\mnras{{MNRAS}}%
\newcommand  \kms      {\ifmmode {\rm km\,s}^{-1} \else km\,s$^{-1}$\fi}
\newcommand  \cc       {\hbox{cm$^{-3}$}}
\newcommand  \cmii     {\hbox{cm$^{-2}$}}
\newcommand  \ergs     {\ifmmode {\rm erg\,s}^{-1} \else erg s$^{-1}$\fi}
\newcommand  \ergcms   {\ifmmode {\rm erg\,cm}^{-2}\,{\rm s}^{-1} \else erg\,cm$^{-2}$\,s$^{-1}$\fi}
\newcommand  \ergcmsA  {\ifmmode{\rm erg\,cm}^{-2}\,{\rm s}^{-1}\,{\rm\AA}^{-1} \else erg\,cm$^{-2}$\,s$^{-1}$\,\AA$^{-1}$\fi}
\newcommand  \ergcmsHz {\ifmmode{\rm erg\,cm}^{-2}\,{\rm s}^{-1}\,{\rm Hz}^{-1} \else erg\,cm$^{-2}$\,s$^{-1}$\,Hz$^{-1}$\fi}
\newcommand  \phcms    {\ifmmode {\rm ph\,cm}^{-2}\,{\rm s}^{-1} \else ,ph\,cm$^{-2}$\,s$^{-1}$\fi}
\newcommand  \phcmsA   {\ifmmode {\rm ph\,cm}^{-2}\,{\rm s}^{-1}\,{\rm\AA}^{-1} \else ph\,cm$^{-2}$\,s$^{-1}$\,\AA$^{-1}$\fi}
\newcommand  \mbh      {\ifmmode M_{\rm BH} \else $M_{\rm BH}$\fi}

\def\mic{\ifmmode \mu{\rm m} \else $\mu$m\fi}
\def\kms{\ifmmode {\rm km\,s}^{-1} \else km\,s$^{-1}$\fi}
\def\Hubble{\ifmmode {\rm km\,s}^{-1}\,{\rm Mpc}^{-1} \else km\,s$^{-1}$\,Mpc$^{-1}$\fi}
\def\ergsec{\ifmmode {\rm ergs\;s}^{-1} \else ergs s$^{-1}$\fi}
\def\ergscm{\ifmmode {\rm ergs\,s}^{-1}\,{\rm cm}^{-2} \else ergs\,s$^{-1}$\,cm$^{-2}$\fi}
\def\ergscmA{\ifmmode {\rm ergs\,s}^{-1}\,{\rm cm}^{-2}\,{\rm \AA}^{-1} \else ergs\,s$^{-1}$\,cm$^{-2}$\,\AA$^{-1}$\fi}
\def\ergscmHz{\ifmmode {\rm ergs\,s}^{-1}\,{\rm cm}^{-2}\,{\rm Hz}^{-1} \else ergs\,s$^{-1}$\,cm$^{-2}$\,Hz$^{-1}$\fi}
\def\Msun{\ifmmode M_{\odot} \else $M_{\odot}$\fi}
\def\Lsun{\ifmmode L_{\odot} \else $L_{\odot}$\fi}
\def\Zsun{\ifmmode Z_{\odot} \else $Z_{\odot}$\fi}
\def\qo{\ifmmode q_{0} \else $q_{0}$\fi}
\def\Ho{\ifmmode H_{0} \else $H_{0}$\fi}
\def\ho{\ifmmode h_{0} \else $h_{0}$\fi}
\def\qo{\ifmmode q_{0} \else $q_{0}$\fi}
\def\ao{\ifmmode a_{0} \else $a_{0}$\fi}
\def\to{\ifmmode t_{0} \else $t_{0}$\fi}
\def\gtsim{\raisebox{-.5ex}{$\;\stackrel{>}{\sim}\;$}}
\def\Halpha{\ifmmode {\rm H}\alpha \else H$\alpha$\fi}
\def\Hbeta{\ifmmode {\rm H}\beta \else H$\beta$\fi}
\def\hb{\ifmmode {\rm H}\beta \else H$\beta$\fi}
\def\Hgamma{\ifmmode {\rm H}\gamma \else H$\gamma$\fi}
\def\Hdelta{\ifmmode {\rm H}\delta \else H$\delta$\fi}
\def\Lya{\ifmmode {\rm Ly}\alpha \else Ly$\alpha$\fi}
\def\Lyb{\ifmmode {\rm Ly}\beta \else Ly$\beta$\fi}
\def\hi{\ifmmode \mbox{{\rm H}\,{\sc i}} \else H\,{\sc i}\fi}

\def\hei{He\,{\sc i}}

\def\ciii{\ifmmode {\rm C}\,{\sc iii} \else C\,{\sc iii}\fi}
\def\civ{C\,{\sc iv}}

\def\mgii{\ifmmode {\rm Mg}{\textsc{ii}} \else Mg\,{\sc ii}\fi}

\def\feii{Fe\,{\sc ii}}

\def\o5007{[O\,{\sc iii}]\,$\lambda5007$}

\def\ne212m {[Ne\,{\sc ii}]\,$12.8 \mu m$}

\newcommand{\LHD} {\ifmmode L_{\rm HD} \else $L_{\rm HD}$\fi}
\newcommand{\CFHD}{\ifmmode {\rm CF}_{\rm HD} \else ${\rm CF}_{\rm HD}$\fi}
\newcommand{\Lbol} {\ifmmode L_{\rm bol} \else $L_{\rm bol}$\fi}
\newcommand{\Lop} {\ifmmode L_{\rm 5100} \else $L_{\rm 5100}$\fi}
\def \Lir{$L_{\rm IR}$}
\def \Ledd{$L/L_{\rm Edd}$}

\newcommand{\RHD} {\ifmmode R_{\rm HD} \else $R_{\rm HD}$\fi}
\newcommand{\RBLR} {\ifmmode R_{\rm BLR}(\hb) \else $R_{\rm BLR}(\hb)$\fi}

\def  \RNLR        {\hbox{$ {R_{\rm NLR}} $}}

\def  \kms         {\hbox{km s$^{-1}$}}          
\def  \cc          {\hbox{cm$^{-3}$}}

\def  \cmii        {\hbox{cm$^{-2}$}}
\def  \La          {\ifmmode {\rm Ly}\alpha \else Ly$\alpha$\fi}
\def  \Ka          {\ifmmode {\rm K}\alpha \else K$\alpha$\fi}
\def  \Lb          {\ifmmode {\rm L}\beta \else L$\beta$\fi}
\def  \Ha          {\ifmmode {\rm H}\alpha \else H$\alpha$\fi}
\def  \Hb          {\ifmmode {\rm H}\beta \else H$\beta$\fi}
\def  \HeINIR    {\ifmmode {\rm He}\,{\sc i}\,\lambda10830 \else He\,{\sc i}\,$\lambda10830$\fi}
\def  \Pa          {\ifmmode {\rm P}\alpha \else P$\alpha$\fi}
\def  \CIIIb       {\ifmmode {\rm C}\,{\sc iii]}\,\lambda1909 \else C\,{\sc iii]}\,$\lambda1909$\fi}
\def  \CIV         {\ifmmode {\rm C}\,{\sc iv}\,\lambda1549 \else C\,{\sc iv}\,$\lambda1549$\fi}
\def  \MgII         {\ifmmode {\rm Mg}\,{\sc ii}\,\lambda2798 \else Mg\,{\sc ii}\,$\lambda2798$\fi}
\def  \mgii         {\ifmmode {\rm Mg}\,{\sc ii} \else Mg\,{\sc ii}\fi}
\def  \OVI         {\ifmmode {\rm O}\,{\sc vi}\,\lambda1035 \else O\,{\sc vi}\,$\lambda1035$\fi}

\def \spitzer      {{\it Spitzer}}

\def \iras      {{\it IRAS}}

\def\tv{\ifmmode \tau_V \else $\tau_V$}
\def\Chisq{\ifmmode \chi^{2} \else $\chi^{2}$}

\def \lowSN {15}
\def \woNIR {37}
\def \wNIR {78}
\def \SBdom {15}
\def \SBdomNLS {12}
\def \SBdomBLS {3}

\title[Graphite Dust in AGN]{Hot Graphite Dust and the Infrared Spectral Energy Distribution of Active Galactic Nuclei}

\author[Mor \&\, Netzer]
{Rivay Mor $^1$\thanks{E-mail: rivay@wise.tau.ac.il} and Hagai~Netzer $^{1}$ \\
$^1$School of Physics and Astronomy and the Wise Observatory, The Raymond and Beverly Sackler
 Faculty of Exact Sciences, Tel-Aviv University, Tel-Aviv 69978, Israel\\}

\date{Accepted 2011 October 21. Received 2011 September 26}

\pagerange{\pageref{firstpage}--\pageref{lastpage}} \pubyear{2011}

\begin{document}

\label{firstpage}

\maketitle

\begin{abstract}
\noindent
We present a detailed investigation of the near-to-far infrared (IR) spectral
energy distribution (SED) of a large sample of Spitzer-observed active galactic nuclei (AGN).
We fitted the spectra of 51 narrow line Seyfert 1 galaxies (NLS1s) and 64 broad line Seyfert 1
galaxies (BLS1s) using a three component model: a clumpy torus, a dusty narrow line region (NLR) and 
hot pure-graphite dust located in the outer part of the broad line region (BLR).
The fitting is performed on star formation (SF) subtracted SEDs using SF templates 
that take into account the entire range of possible host galaxy properties. 
We find that the mid-IR intrinsic emission of NLS1s and BLS1s are very similar, regardless of the AGN luminosity,
with long wavelength downturn at around 20--25 \mic.
We present a detailed model of the hot dust component that takes into account the distribution
of dust temperature within the clouds and their emission line spectrum. 
The hot dust continuum provides a very good fit to the observed near-IR continuum spectrum. 
Most line emission in this component is dramatically suppressed, except \MgII\ and \hei\ lines that are still
contributing significantly to the total BLR spectrum. 
We calculate the covering factors of all the AGN components and show that the
covering factor of the hot-dust clouds is about 0.15-0.35, similar to the covering factor of the torus,
and is anti-correlated with the source luminosity and the normalized accretion rate.
\end{abstract}


\section{Introduction}
\label{sec_intro}

The unification scheme of active galactic nuclei (AGN) requires an anisotropic obscuring structure
that surrounds the central accreting black hole \cite[e.g.,][]{Krolik1988,Antonucci1993}.
In this picture, the bulk of the radiation from the central engine is absorbed by the obscuring
structure and re-emitted mainly at mid-infrared (MIR) wavelengths. 
The infrared (IR) spectral energy distribution (SED) of AGN also shows emission by star forming (SF) regions in the host galaxy.
Recent studies have shown that polycyclic aromatic hydrocarbon (PAH) features
can serve as good indicators for this SF activity \cite[e.g.,][]{Schweitzer2006, LaMassa2010a}.
Measuring and subtracting the SF contribution to the IR SED enables to focus on the AGN-produced SED.
\cite{Netzer+07_QUEST2} used the PAH features to subtract the SF contribution from the average
SED of the QUEST sample of PG quasars (QSOs). This was done for two sub groups of low and high far-IR (FIR)
luminosity. The two subtracted SEDs were found to be remarkably similar suggesting that most AGN have
similar \textit{intrinsic} SEDs. \cite{Deo2009} reached a similar conclusion while investigating samples of both 
type-I and type-II Seyfert galaxies. Both studies found that the intrinsic AGN SEDs exhibit two main bumps
around $\sim10$ and $20$-\mic. These are associated with broad silicate emission features as well as emission from cooler dust clouds.
More recently \cite{Mullaney2011} found that the second bump may peak at somewhat longer wavelengths ($\sim30$-\mic)
indicating the presence of even colder dust.

A main component of the obscuring circumnuclear structure is believed to be a dusty torus.
The MIR SED of such a torus depends on its dimensions and geometry,
the density distribution and the dust grain properties.
Several attempts have been made to model such tori assuming smooth density distributions
\cite[e.g.,][]{Pier1992, Pier1993, Granato1994, Efstathiou1995, vanBemmel2003, Schartmann+05}
and clumpy density distributions \cite[e.g.,][]{Nenkova+08a,Honig2010}.
Other AGN components can contribute to the observed MIR spectrum.
Some of this emission may originate farther from the central radiation source,
at distances exceeding the dimensions of the torus.
Dusty gas in the narrow-line region (NLR) may be the source of such radiation
\cite[hereafter M09]{Schweitzer2008,Mor2009}.
Thus, a significant contribution at $\gtsim$20-\mic\ due to components not related to the torus,
must be considered.

Dust reverberation measurements of several nearby AGN, based on the V and K band emission,
lead to the conclusion that the near IR (NIR) emission in these sources is dominated
by thermal radiation from hot dust very close to the centre \cite[][]{Minezaki+04,Suganuma+06}.
There have been several studies that fitted the NIR to MIR SEDs of AGN using a black-body spectrum
to represent emission from such hot dust (e.g., \citealt{Edelson&Malkan86, Barvainis87, Kishimoto+07,Riffel2009}; M09; \citealt{Deo2011}).
More recently, \cite{Landt2011a} found similar results by fitting the optical to NIR SED of 23 AGN.
The modelled temperature of this component, in all these studies, is high ($\gtsim1200\,\rm{K}$),
regardless of the AGN luminosity, and is consistent with pure-graphite dust emission (M09).
\cite{Mor2011} used the (observed) optical to MIR SEDs of $\sim15000$ type-I AGN 
to conclude that a hot pure-graphite dust component is observed in the vast majority of such sources.
Several studies found that the NIR luminosity strongly correlates with the luminosity of the AGN \cite[e.g.,][]{Gallagher2007, Mor2011}.
\cite{Gallagher2007} concluded that this tight relation suggests a constant covering factor of dusty clouds
regardless of AGN luminosity. 
\cite{Mor2011} found that the covering factor of the hot dust component decreases with increasing AGN bolometric luminosity.

Following M09, we aim to investigate a larger sample of type-I AGN with a broad range of 
luminosities, black hole masses (\mbh), and normalized accretion rates (\Ledd).
We fit the observed MIR spectra of 115 type-I AGN using a 3-component model
made of a clumpy torus, dusty NLR clouds and very hot dust clouds.
Our main goal is to explore the physical properties of these components, focusing mainly on the hot dust.
The high quality spectra made available by the IRS spectrometer on-board
the \textit{Spitzer} Space Observatory \citep{Houck+04} allows us to explore these ideas.
In \S\ref{sec_sample} we describe the observational data.
In \S\ref{sec_spec_decomp} we detail our model and the fitting procedure
and in \S\ref{sec_results} we present the results of this procedure and discuss their implications.


\section{Sample Selection Observations and Data Reduction}
 \label{sec_sample}
%

Our database consists of local type-I AGN collected from several samples, covering a wide range in luminosity and
divided into two groups of NLS1s and BLS1s, according to their FWHM[\Hbeta], with a division line at 2000 \kms.
20 NLS1s in our sample are taken from the \textit{ROSAT} sample of nearby AGN \cite[]{Thomas1998}, and are a part of
the \spitzer\ PID 20241 (PI D. Lutz) recently described by \citet[hereafter S10]{Sani2010}. 
S10 supplemented these objects by a large number of additional \spitzer/IRS archival observations of NLS1s and BLS1s.
These additional objects were selected from the 12th edition of the Catalog of Quasars and Active Nuclei compiled by \cite{Veron-Cetty&Veron_06}. 
We also added to the sample all AGN in the QUEST \spitzer\ spectroscopy project (PID 3187, PI S. Veilleux), 
which is described in detail in \cite{Schweitzer2006} , \cite{Netzer+07_QUEST2} and M09. 
Most of the QUEST objects are Palomar-Green (PG) QSOs \cite[]{Schmidt&Green83} and are taken from \cite{Guyon+06}.
In total, our sample consists of 115 sources (51 NLS1s, 64 BLS1s) spanning a luminosity range of
\Lop$\approx10^{42.2-45.9}\,\ergs$ where \Lop\ stands for $\lambda L_{\lambda}$ at rest wavelength 5100~\AA.

The \spitzer\ observations and the data reduction procedure of the S10 and QUEST samples are detailed in S10 and
\cite{Schweitzer2006}, respectively and are summarized here for completion.
IRS spectra for all objects were taken in either low or high resolution modes and cover the wavelength range of 5--35 \mic\ in the observed frame.
The standard slit widths of 3\arcsec.6 to 11\arcsec.1 include flux from the host galaxies and the vicinity of the AGN.
We started the data reduction from the basic calibrated data (BCD) provided by the \spitzer\ pipeline.
Specially developed IDL-based tools are then used for removing outlying values for individual pixels and for sky subtraction. 
The SMART tool \cite[]{Higdon+04} is used for extraction of the final spectra.

We supplemented the \spitzer\ spectra with NIR data obtained from the 2MASS 
extended and point source catalogues \cite[]{Skrutskie+06_2mass}.
Approximately 2/3 of the sources have been detected by the 2MASS survey.
We re-measured the 2MASS images using an IRS-slit-like aperture in order to avoid biases 
regarding the large aperture used in the 2MASS catalogues. 
This gives a better calibration to the NIR measurements with respect to the spectrum of the source, 
and is more relevant for the extended sources in our sample (~50\% of the sample).
The main source of uncertainty related to the NIR data is due to the fact that many
of these images were taken more than 20 years before the \spitzer\ observations.
For the longer wavelengths of the SED we use \iras\ 60-\mic\ photometry \cite[]{Neugebauer+84_iras}.
Aperture corrections, similar to those done for the 2MASS data, are not possible for the \iras\ data due to 
the low spatial resolution of the \iras\ instrument ($\sim2$\arcmin). 
This contributes to the uncertainty in the FIR flux of the extended objects in our sample and will be discussed in
\S\ref{subsec_spec_decomp_SF_host}.

In order to compute \Lbol\ for all sources we adopt the \cite{Marconi2004} ``intrinsic'',
luminosity dependant SED, which provides a polynomial prescription for estimating $\lambda L_{\lambda}$(4400-\AA) at every \Lbol.
We use this prescription for \Lop\ by adopting a $f_{\nu}\propto \nu^{-0.5}$ power-law approximation.
For the S10 sample, values of \Lop\ were obtained from various works (see S10 and references therein)
For most objects in the QUEST sample the values of \Lop\ are obtained from the observations
of \cite{Boroson&Green92}. For PG~1244+026, PG~1001+054, and PG~0157+001 we used spectra from the
seventh data release of the Sloan Digital Sky Survey \cite[SDSS/DR7,][]{Abazajian2009a}
that were measured in the way described in \cite{Netzer&Trakhtenbrot07}.
We have estimated \mbh\ and \Ledd\ for all sources using the procedure described in 
Netzer \& Trakhtenbrot (2007). In this procedure (the ``virial" mass determination, see Eq.~1 in Netzer \& Trakhtenbrot 2007),
\Lop\ and FWHM(\Hb) are combined to obtain \mbh. 
\Ledd\ is obtained by using the adopted bolometric correction factor (BC).

There are two uncertainties associated with the use of \Lbol.
The first is source variability, which is an important
effect since the optical and MIR observations were separated by many years. We estimate this uncertainty
to be a factor of $\sim1.5$. The second uncertainty involves the approximation used for the BC.
We estimate this uncertainty to be $\sim$30\%.
Both uncertainties affect the derived model parameters such as the covering factors and the distances to the NLR and hot dust clouds.
Because of this, we do not attach great importance to specific values obtained from our best acceptable models.
On the other hand, the sample is large enough to enable a significant analysis
of its mean properties since the larger of these effects, due to source variability, is likely to be random.
The uncertainty due to the estimate of BC is smaller but more problematic since
the expression we use can under-estimate or over-estimate \Lbol\, for {\it all sources}.
This can introduce systematic differences, e.g. a smaller covering factors for all sources.


\section{Spectral Decomposition}
\label{sec_spec_decomp}
One of the main goals of the present work is to investigate the 2--60 \mic\ SED of the sources in our sample, 
following the general scheme of M09. 
Our models include different physical components: a very hot pure-graphite dust, 
a dusty torus containing graphite and silicate grains, a dusty NLR, and a SF host galaxy.
We start by separating the AGN-related components from the host galaxy component.
The subtraction of the SF contribution is a major challenge and a major source of uncertainty
mainly at longer wavelengths where the SF is dominant.
The adopted procedure is described below.

\subsection{Star Forming Regions in the Host Galaxy}
\label{subsec_spec_decomp_SF_host}

\begin{figure}
\includegraphics[width=9cm, height=9cm]{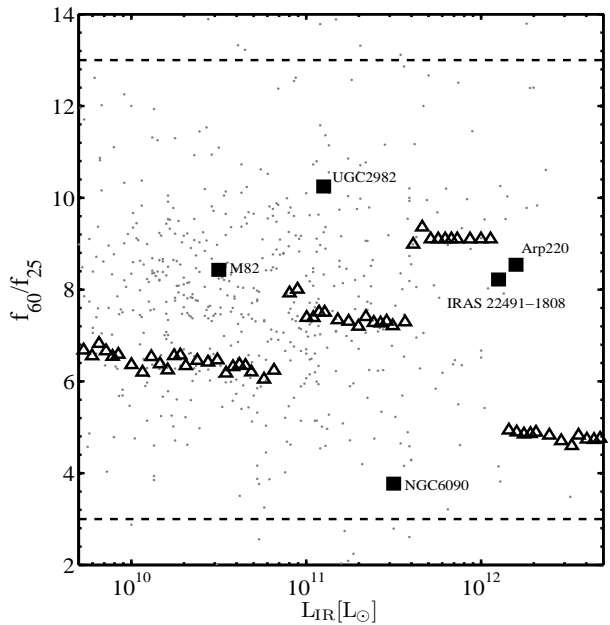}
\caption{Ratio between FIR and MIR fluxes vs. IR luminosity.
Gray points represent IR bright galaxies from the The IRAS Revised Bright Galaxy Sample \citet[]{Sanders+03}.
Calculated ratios from theoretical SF templates of \citet{Chary&Elbaz01} are presented as triangles
and the \textit{observed} ratio in the 5 galaxies used as SF templates in this paper are shown as black squares.
The horizontal dashed lines show flux ratios of 3 and 13.
}
\label{fig:f60-f25_ratio}
\end{figure}

\begin{figure}
\includegraphics[width=9cm, height=9cm]{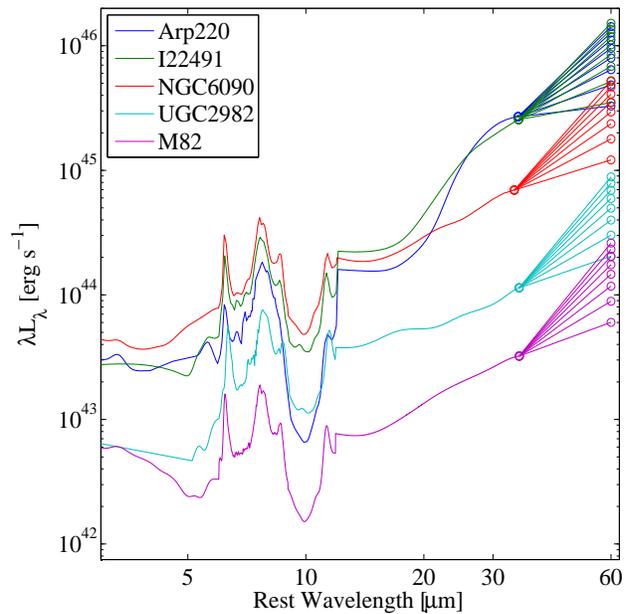}
\caption{\spitzer/IRS spectra of 5 SF galaxies that were used to construct the SF templates.
Each spectrum is supplemented with 8 simulated data points at 60 \mic\
to sample the observed range in $f_{60}/f_{25}$ shown in Fig.~\ref{fig:f60-f25_ratio}}
\label{fig:SB_templates}
\end{figure}

In order to properly account for the SF emission we need SF templates that represent a large range of properties of SF galaxies.
We used \spitzer/IRS spectra of 5 different SF galaxies: M82, UGC~2982, NGC~6090, IRAS~22491-1808, and Arp~220.
These galaxies span a wide range of IR luminosity (\Lir$\approx10^{10.5-12.2}\,\Lsun$) and
PAH luminosity (e.g., $L_{PAH\,7.7\mic}\approx10^{8.4-9.8}\,\Lsun$).
Most important, we need to take into account the large range and large scatter in the ratio between the FIR and MIR luminosities,
known to exist in SF galaxies \cite[e.g.,][]{Sanders+03}.
As can be seen in Figure~\ref{fig:f60-f25_ratio}, the scatter in $f_{60}/f_{25}$ (the flux ratio) is between $\sim$3-13.
This scatter introduces an uncertainty in the determination of the intrinsic AGN SED for a specific object
(see further discussion in \S\ref{subsec_res_SF_intrinsic_AGN}). 
For each of our 5 SF galaxies we choose 8 values within this range (3-13) and simulate a 60 \mic\ data point accordingly.
We use the simulated 60 \mic\ data points to construct, from each MIR spectrum, 8 SF templates covering the NIR to FIR wavelength range.
In total, our set of SF templates consists of 40 (5$\times$8) templates covering a large range of IR luminosity, FIR/MIR ratio, 
and PAH intensities. 
Finally, the observed spectra of the templates are smoothed to avoid any systematic features in the subtracted AGN spectra.
The 40 SF templates are shown in Figure~\ref{fig:SB_templates}.

We subtract each template from the observed NIR-NIR SEDs using the intensity of the PAH emission features.
The PAH emission is assumed to be solely due to SF in the host galaxy.
The 7.7-\mic\ PAH feature was detected in 50 sources in our sample.
This feature is strong and can be used as a reliable indicator for SF activity.
Recent studies have shown that the ratio between the 11.3 and 7.7-\mic\ features in AGN
may differ significantly from those observed in SF galaxies \cite[]{Diamond-Stanic&Rieke10,Sales+10}.
To further test this issue we compared the luminosity ratio of 
the 11.3 and 7.7-\mic\ features in our sample to our SF templates.
All our sources exhibit a much larger ratio than observed in the templates, suggesting that either the 11.3 \mic\
feature is stronger in AGN compared with SF galaxies at similar IR luminosities,
or that the AGN activity may suppress the shorter wavelength feature.
The measurement of both features is problematic. The 7.7-\mic\ feature is 
often blended with the adjacent 8.6-\mic\ feature, and the 11.3-\mic\ feature is strongly affected by 
the adjacent 10-\mic\ broad silicate feature.

For each SF template we choose a normalization factor and require that after the subtraction the flux at
the wavelengths corresponding to the PAHs will not exceed the noise level.
Application of this criterion to the 11.3-\mic\ feature resulted in deep 
``absorption'' features in the subtracted spectra at the wavelengths of all the other PAH features (6.2, 7.7, and 8.6-\mic).
Applying the criterion to the 7.7-\mic\ feature resulted in residual flux only in the 11.3-\mic\ wavelength range.
We therefore choose to apply this criterion to the 7.7-\mic\ feature.
This can contribute to the uncertainty in the determination of the SF contribution and 
the shape of the intrinsic AGN SED. We further discuss this issue in \S\ref{subsec_res_SF_intrinsic_AGN}
For objects with no PAH detection the subtraction is limited according to the level of noise in the
relevant wavelength ranges.

Given the large scatter in $f_{60}/f_{25}$, subtraction of different templates using the same
criterion at short wavelengths may imply very different contribution of the SF component to the FIR emission.
Given our assumption that the FIR emission in AGN is dominated by SF in the host galaxy, 
we prefer subtractions in which the SF contribution to the 60-\mic\ flux is above 80\%. 
The combination of all templates that satisfy both criteria at short and long wavelengths define the 
acceptable range of SEDs. As explained in \S\ref{sec_sample}, there is another source of uncertainty
related to the low spatial resolution of the \iras\ observations. This uncertainty propagates into the determination of the SF component
and as a consequence affects the determination of the intrinsic AGN SED
(see \S\ref{subsec_res_SF_intrinsic_AGN} and \S\ref{subsec_dis_NLR} for more discussion).

We calculate the SF contribution to the MIR luminosity by integrating over the scaled SF template between 2--35 \mic.
The median contribution is 20\% with a wide distribution between 0 and 75\%.
\SBdom\ sources in our sample exhibit a SF dominated MIR SED, i.e. 10 \mic\ silicate feature in 
absorption and very strong PAH emission features \cite[see also][]{Deo2009}. 
Indeed, for these sources the SF contribution is much higher compared with the rest of the sample. 
Moreover, SF dominated SEDs are more common among NLS1s (\SBdomNLS\ sources)
compared with BLS1s (\SBdomBLS\ sources). This is consistent with the findings of S10 that NLS1s often 
exhibit much stronger SF activity compared with BLS1s for the same AGN luminosity.
SF dominated MIR SED may also be the result of weak AGN contribution to this region in the SED due to a small covering factor.
We return to this issue in \S\ref{subsec_dis_covering_factors}.

Subtraction of SF templates with large MIR contribution ($\gtrsim35\%$) is problematic.
In these cases the SF contribution, in certain narrower wavelengths ranges (e.g. 6.5--7.5 \mic), can 
exceed the observed flux, thus the remaining AGN SED becomes meaningless. The fact that the SF template 
can exceed the observed flux at certain wavelengths while the integrated flux over the entire 2--35 \mic\
range can be as low as 35\%, is related to the different SED shapes and PAH intensities in SF galaxies.
We omit these sources from the following analysis of the intrinsic AGN SED.
The subtraction of the SF template has profound effect at wavelengths longer than $\sim5$-\mic,
and very little effect on the hot dust component.

\subsection{Clumpy Torus with Silicate-type Dust}
\label{subsec_spec_decomp_torus}

The first AGN-related component is a dusty torus surrounding the central energy source.
We use the clumpy torus models of \cite{Nenkova+08a} in a similar way to that described in M09.
The model accepts \Lbol\ as an input to calculate the normalization of the SED.
Unfortunately, the normalizations that were calculated before September 2010, and used in M09,
were found to be off by a factor of $\sim$2--3 \cite[see erratum by][]{Nenkova+08erratum}. 
By repeating \textit{exactly} the M09 fitting procedure for the current sample, and using the corrected normalization factors,
we could not fit the spectra with low inclination angle models, i.e. those representing type-I AGN.
Such corrected models emit more flux than actually observed.
This suggests that in these cases the torus is exposed to too much incoming radiation.

A possible solution to the above problem is related to the hot dust component found in M09. 
Such a component, located between the central source and the inner edge of the torus, 
must affect the radiation transfer and the energy balance in the system.
This component can have a relatively large covering factor causing a significant 
fraction of the incoming radiation to be absorbed and re-emitted at NIR wavelengths. 
This component is different from the ''standard`` clumpy torus clouds since 
the temperature of its dust is too high for silicate grains to survive 
(see \S\ref{subsec_dis_hot_dust}).
Thus, the clumpy torus model must be amended to include a pure-graphite dust component 
in order to provide a full solution to the observed spectrum.

To overcome the difficulty associated with the incomplete torus model we
added an additional free parameter to the fitting procedure which in practice is a free normalization parameter. 
This parameter represents the fraction of the total radiation reaching the clumpy torus
i.e., the fraction that is not absorbed by the hot pure-graphite dust component.
The new procedure does not provide a full solution since it does not 
include the different SED shapes ``seen'' by the torus, but only accounts for the total energy.
It provides an effective way to determine the contribution of the 
torus component to the IR SED (the torus covering factor).

The main parameters and assumptions of the clumpy torus model are detailed in \cite{Nenkova+08a} and M09.
Unfortunately, the additional free normalization of the torus makes it impossible to 
fully constrain the geometrical parameters of the clumpy torus.
However, the contribution of the torus to the MIR emission can still be reliably determined. 
The covering factor of the torus, $CF_{Torus}$, can be deduced from the ratio between its total
luminosity and the bolometric luminosity of the central source.
This definition of the covering factor is equivalent to $f(i)$ in M09 (eq.~5 there).
 
\subsection{Dusty NLR}
\label{subsec_spec_decomp_NLR}

The second component of the model represents a collection of dusty NLR clouds.
The motivation for this component is explained in \cite{Schweitzer2008} and in
M09 where it was shown that such a component can contribute significantly to the 
MIR flux of luminous AGN.
The properties assumed here for these clouds are similar to the ones used in M09.
We assume constant column density clouds with $N_{\rm H}=10^{21.5}\,\cmii$.
We further assume constant hydrogen density of 10$^{5}$ \cc, solar composition,
galactic dust-to-gas ratio, and ISM depletion. The important physical parameters for this component are
the cloud-central source distance (which determines the dust temperature), the incident SED
and the dust column density. 
For more information on this component see M09.

\subsection{Hot Pure-Graphite Dust}
\label{subsec_spec_decomp_hot_dust}

The third component represents a collection of dusty clouds of gas located at the inner edge
of the torus. The motivation for this component is explained in M09 where it was shown that 
such a component is necessary to explain the NIR emission of type-I QSOs.
\cite{Mor2011} have shown that a hot dust component is present in $\gtrsim80$\% of type-I AGN and is significantly luminous.
The pure-graphite dust must be external to the broad-line region (BLR), where dust cannot survive, 
and internal to the ``standard'' clumpy torus, where the distances are large enough to allow silicate-type grains.

The physical scale of the hot-dust region is set by two different conditions.
The outer boundary is the sublimation radius appropriate for a ``typical'' dust
composed of both (average size) silicate and graphite grains.
This is given by
\begin{equation}
R_{d,Si} \simeq 1.3 \times \left (\frac{L_{\rm bol}}{10^{46}\,\ergs} \right )^{1/2}  \left (\frac{1500\,\rm{K}}{T_{\rm sub}}\right )^{2.6} \, \rm{pc}.
\label{eq:R_d_Si}
\end{equation}
The inner boundary is the sublimation radius of pure-graphite grains
with a sublimation temperature of 1800 K,
\begin{equation}
R_{d,C} \simeq 0.5  \left (\frac{L_{\rm bol}}{10^{46}\,\ergs} \right )^{1/2} \left (\frac{1800\,\rm{K}}{T_{\rm sub}}\right )^{2.8} \,\rm{pc}.
\label{eq:R_d_c}
\end{equation}
This is the innermost radius where an averaged-size graphite dust grain can survive.
Here we suggest that the pure-graphite dust is located in BLR clouds that are between the silicate and graphite
sublimation radii. We note that these radii depend on the assumed grain properties, mostly their size,
and are not to be taken at face value but rather as typical scales of the system.

We investigate the conditions in this region by calculating photoionization models that extend all the
way from the innermost BLR to the dusty torus. 
The numerical code used for the calculation of the IR continuum emitted by the clouds,
as well as their emission line spectrum, is the 2009 version of the code ION which is described in 
detail in various earlier publications \cite[e.g.,][and references therein]{Netzer06}. 
The code includes all the relevant atomic and dust-related processes and was compared 
with CLOUDY (by G. Ferland) to ensure similar results under a large range of conditions including dusty and
dust-free gas.

We adopted the ``cloud model'' of the BLR which we consider to be the one most relevant to the
situations considered here. This is one of three possible generic BLR models addressed in the literature. 
It includes a system of clouds that preserve their mass as they move in or out. 
The properties of the clouds depend on the radial coordinate $r$ through an external parameter, $s$, that
describes the external confining pressure (presumably magnetic in origin).
The model was originally suggested by \cite{Rees1989} and was extended to more realistic cases in
\cite{Netzer1990} and in \cite{Kaspi1999}.
The main model ingredients are the gas density in the clouds ($N(r) \propto r^{-s}$), 
the cloud size ($R(r) \propto r^{s/3}$), the column density ($N_{col}(r) \propto r^{2s/3}$), 
and the geometrical cross section of the clouds ($A(r) \propto R(r)^2$).
An additional parameter, $p$, specifies the volume density of the clouds,
($n_c(r) \propto r^{-p}$) and thus the radial-dependent covering fraction of the system, $C(r)$. 
Using the above definition we get, $dC(r) \propto r^{2s/3 -p} dr$.
  
The actual calculations assume constant density clouds, the parameters $s$ and $p$, 
and the inner boundary conditions on $N$ and $N_{col}$. These can be adjusted to 
give the best agreement with the observations of a specific AGN. 
\cite{Kaspi1999} investigated a large range of dust-free BLR properties in attempt to fit the time-variable emission
line spectrum of NGC 5548. They found that the values that are consistent with the observations are
$1 \leq s \leq 1.5$ and $ 1 \leq p \leq 2$. 

Two other types of BLR models were not considered in this work.
The first is the ``local optimally-emitting clouds'' model \cite[LOC; e.g.,][]{Baldwin1995, Korista1997},
which assumes a large range of conditions (density, column density etc.) at every location in the BLR.
The second is the wind model \cite[e.g.,][]{Murray1995} which is based on the assumption that BLR clouds condense out
of a fast outflow, which originates very close to the central source, perhaps in the vicinity of the central accretion disk. 
The emission line characteristics of the wind model have never been compared in detail with specific AGN observations. 
Our preference of the cloud model is based in part on several recent observations that directly detect 
cloud-like objects in the general region of the BLR in low luminosity AGN
\cite[see e.g.,][and references therein]{Maiolino2010, Turner2009}

The parameters of the generic model which has been used in this study are:
$s=1$, $p=1$, $N(r=R_{d,C})=10^{9.8}$ cm$^{-3}$  and $N_{col}(r=R_{d,C})=10^{22.66}$ cm$^{-2}$, 
where $R_{d,C}$ is the graphite sublimation radius (eq. \ref{eq:R_d_c}).
This combination gives $U(r=R_{d,C})=1.6\times10^{-2}$ where $U$ is the ionization parameter.
The model further assumes gas composition of 2 \Zsun\ and ISM-type depletion.
The incoming SED is typical of high luminosity AGN and is made of a combination of a central accretion disk
and a high energy X-ray power-law with a break at 50 keV.
The results of the calculations depend mostly on the value of $L_{bol}$ and not on the exact SED shape.

\begin{figure}
\includegraphics[width=9cm, height=9cm]{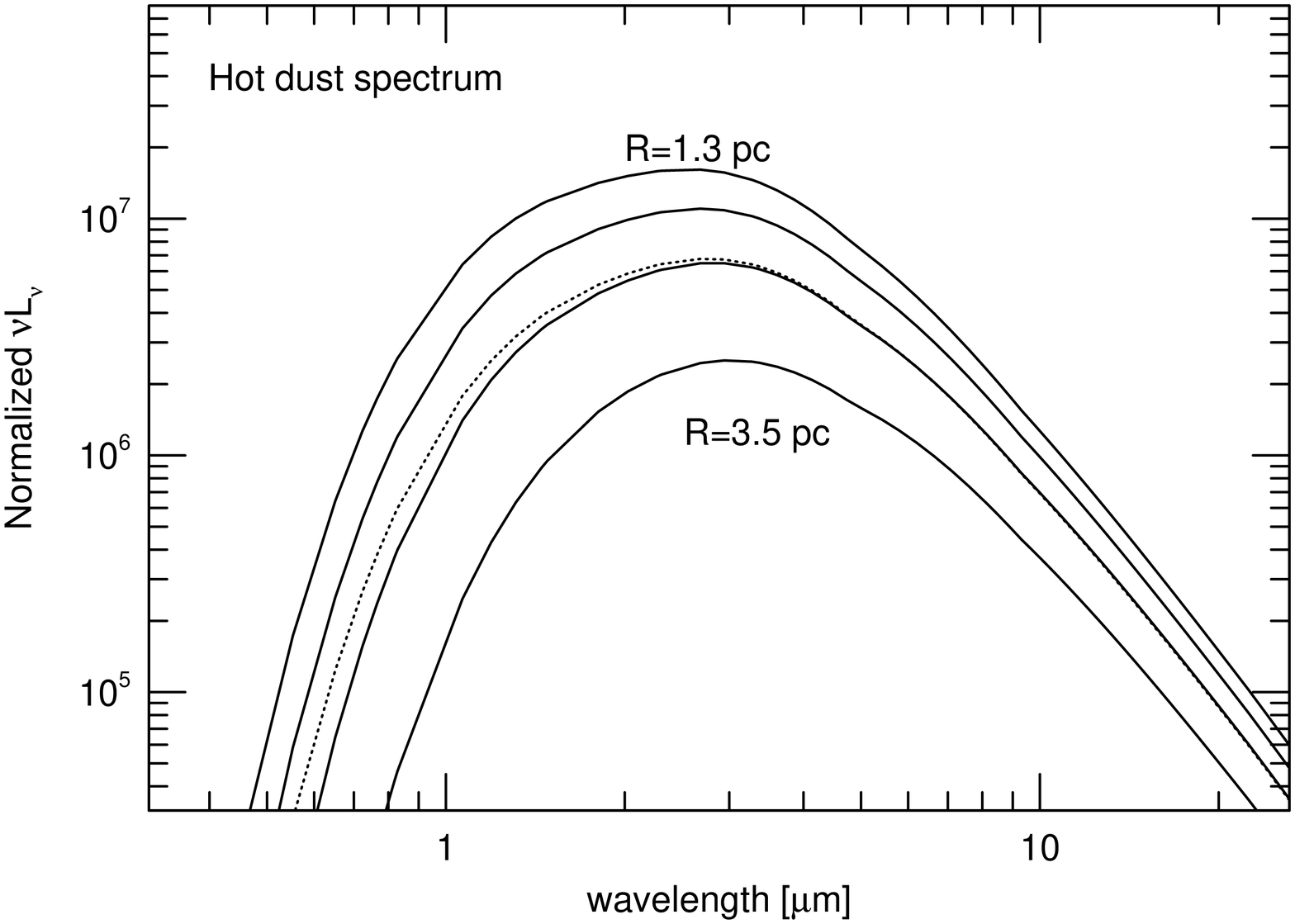}
\caption{Normalized spectra, per unit covering factor, of the hot pure-graphite
dust component. The gas composition is 2 \Zsun\ and the cloud properties are given
in the text. All the spectra shown here are calculated for the case of $\Lop=10^{46}\,\ergs$.
The smallest and largest cloud-source distances in the grid are marked next
to the lines. The dotted line is the mean spectrum of all dusty clouds weighted by their covering
factor.}
\label{fig:hot_dust_grid}
\end{figure}

While the local emission of the graphite dust depends only on the grain properties, the dust temperature
inside the cloud varies by a large factor because of the local grain opacity.
Thus a single cloud spectrum looks like a combination of many modified blackbodies. 
The calculations provide a grid of dust-produced SEDs that are used in the fitting of the
hot dust component. Such a grid is shown in figure~\ref{fig:hot_dust_grid}.
The calculations also provide the emission line spectrum of the ensemble of dusty clouds
(see \S\ref{subsubsec_dis_hot_dust_emission_line}).

\begin{figure*}
\begin{center}$
\begin{array}{cc}
\includegraphics[width=9cm, height=9cm]{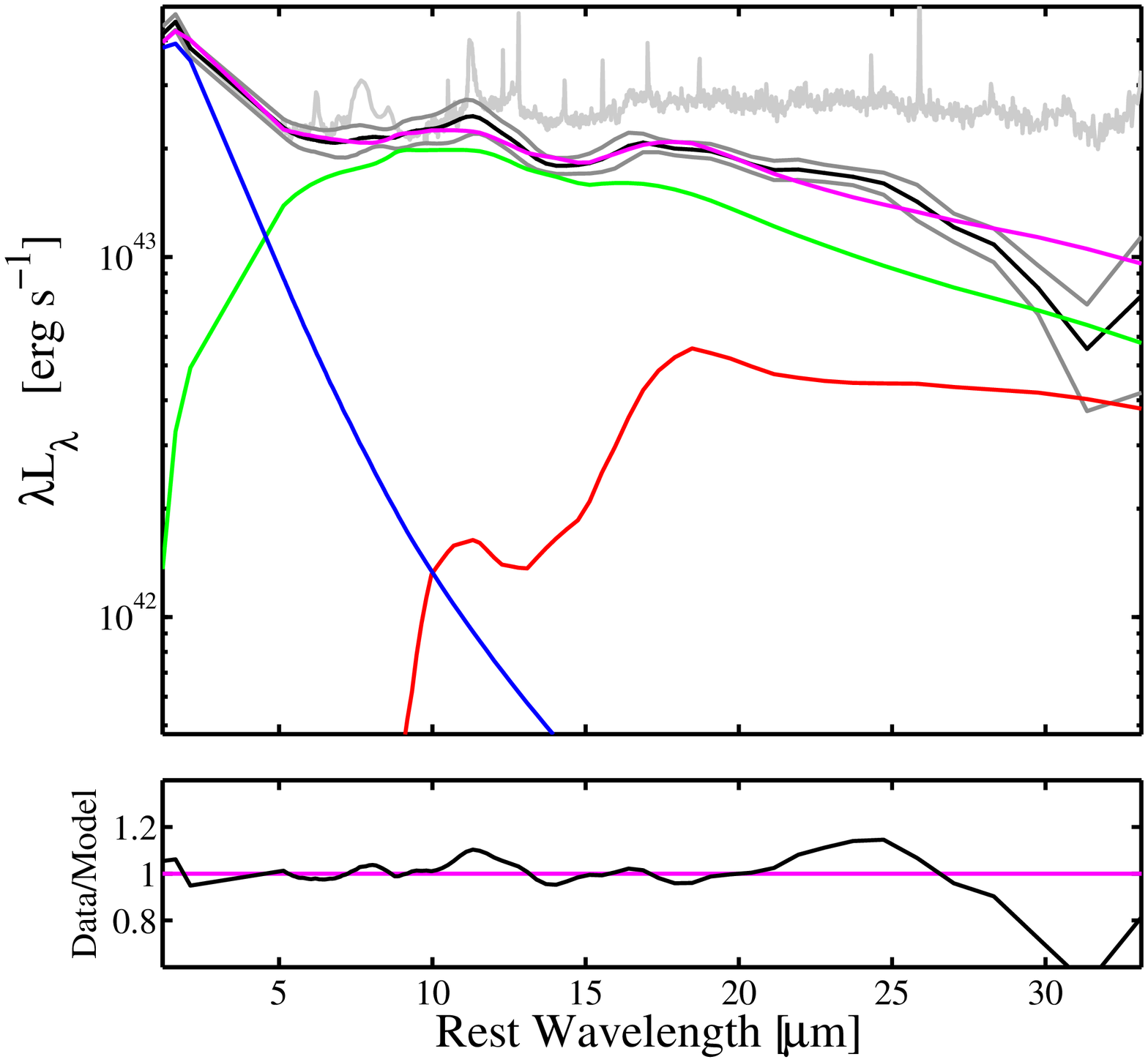} & 
\includegraphics[width=9cm, height=9cm]{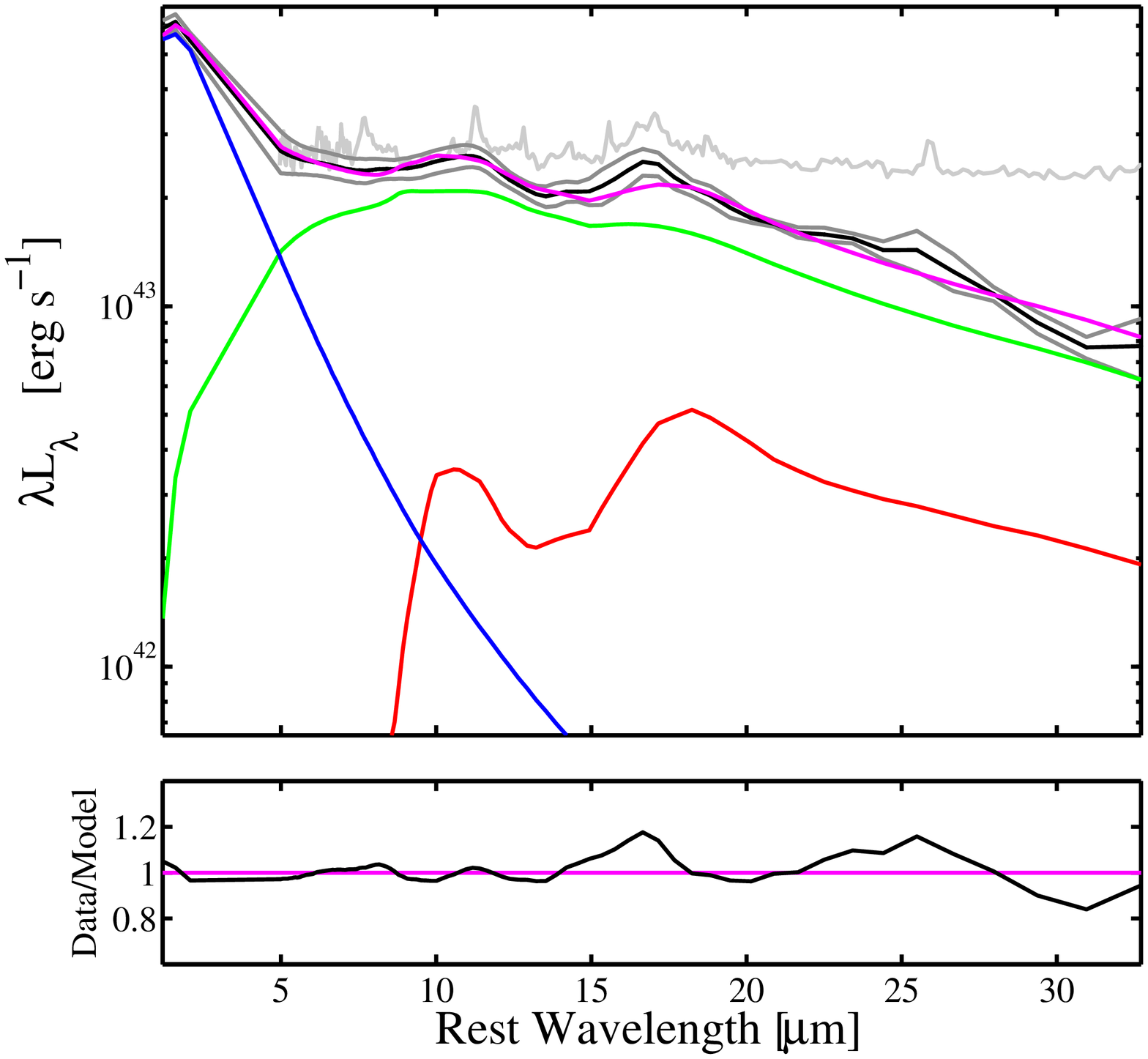}
\end{array}$
\end{center}
\caption{Best fit using the three-component model for two representative cases,
the NLS1 Mrk~896 (left diagram) and the BLS1 Mrk~1146 (right diagram).
Top panels show the best fit model (magenta) and the observed (grey) 
and subtracted and binned (black) spectra. We also show individual components:
torus (green), NLR (red) and hot pure-graphite dust (blue).
In the bottom panels of each diagram we show the quality of the fit in each wavelength bin,
by calculating the ratio between the model and the fitted data.}
\label{fig:fit_example}
\end{figure*}

\subsection{Model Fitting and Fit Quality}
\label{subsec_spec_decomp_fitting_method}

The fitting procedure starts with the SF template subtracted SED that is
assumed to represent the intrinsic AGN continuum.
Although the SF component is subtracted prior to the model fitting,
it introduces another degree of freedom to the procedure.
To fit the SEDs with a 3-component models we use \Chisq\ minimization 
in a similar way to that described in M09. The only difference is that the
normalization of the torus component is now a free parameter.
 
The fitting algorithm computes \Chisq\ values for all possible combinations of torus,
NLR and hot dust models. There are seven free parameters in the torus model,
the normalization parameter and six others that describe its
geometrical properties (\S~\ref{sec_spec_decomp}).
In the NLR model there are two free parameters, the cloud distance and a normalization factor.
The distance is changed in steps of 0.075 dex between 1 and 850 pc
for a source with $L_{\rm bol}=10^{45}\,\ergs$.
For the hot dust, there are also two free parameters similar to the NLR component,
the distance to the cloud and a normalization factor. For example, for a source with $\Lop=10^{46}\,\ergs$
the range of distances is changed in steps of 0.03 dex between 0.4 and 6 pc.

Fig.~\ref{fig:fit_example} shows the best fit models for two representative cases,
the NLS1 Mrk~896 and the BLS1 Mrk~1146.
The top panel of each diagram shows the best fit model (magenta) and the observed (grey) 
and subtracted and binned (black) spectra. We also show individual components:
torus (green), NLR (red) and hot pure-graphite dust (blue).
In the bottom panels of each diagram we show the quality of the fit in each wavelength bin,
by calculating the ratio between the model and the fitted data.


\section{Results and Discussion}
\label{sec_results}

The sample described here was collected from the \spitzer\ archive with no specific selection criteria.
Thus, some sources lack certain data and/or sufficient quality to be proper analysed.
\wNIR\ sources in the sample (38 NLS1s and 40 BLS1s) have NIR photometry. In the following analysis
related to the hot dust component (\S\ref{subsec_dis_hot_dust} and \S\ref{subsec_dis_covering_factors})
we only consider these sources.
The IRS spectra of \lowSN\ sources in our sample have very low signal to noise ratio.
All silicate and PAH features in these sources are undetected and it is impossible to constrain the continuum shape. 
These sources are not included the analysis of the intrinsic AGN SED (\S\ref{subsec_res_SF_intrinsic_AGN}),
the torus SED (\S\ref{subsec_dis_torus}), and the NLR properties (\S\ref{subsec_dis_NLR}).

\subsection{Star Formation in the Host Galaxy and the Intrinsic AGN SED}
\label{subsec_res_SF_intrinsic_AGN}

Our method to construct the SF templates is driven by the intrinsic scatter in the ratio between the 
FIR and MIR fluxes found in SF galaxies. 
The subtraction procedure uses two criteria to determine the best SF template and its normalization. 
The first is the condition that after the subtraction, any remaining flux at the wavelength range of the PAH features
would be consistent with the noise level.
Since our SF library consists of templates that span a large range of $L_{IR}$, only small adjustment 
of the normalization is needed, and the normalization values are usually close to one.
The second criterion is that the SF template would dominate the FIR part (60-\mic) of the SED.

\begin{figure}
\includegraphics[width=9cm, height=9cm]{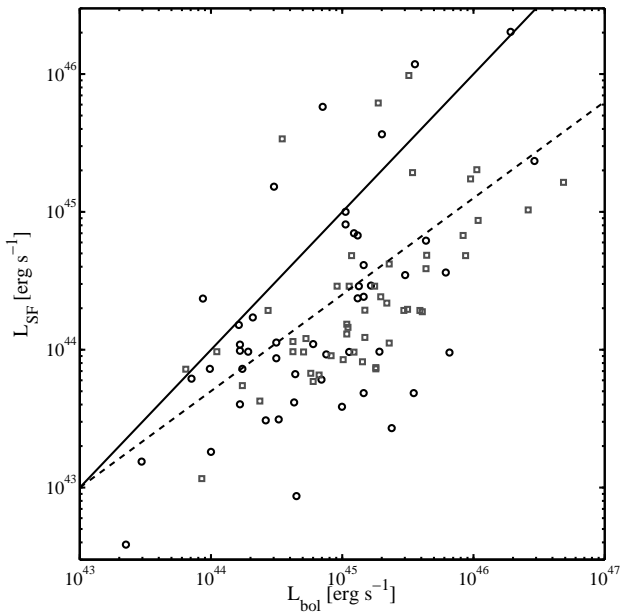}
\caption{Star formation luminosity in AGN host galaxies vs. AGN bolometric luminosity
for the two groups of NLS1s (circles) and BLS1s (squares). 
The SF luminosity is calculated using the IR luminosity of the chosen template and its normalization.
The solid line represents $L_{SF}=\Lbol$ and the dashed line represent the 
relation $L_{SF}=10^{43}\times\left(\Lbol/10^{43}\right)^{0.7} \ergs$ adopted from the correlation shown in \citet{Netzer2009}.}
\label{fig:LSF_Lbol}
\end{figure}

The SF luminosity ($L_{SF}$) of each source is calculated from the subtraction
process using the integrated IR luminosity of the template and its normalization.
Figure~\ref{fig:LSF_Lbol} shows $L_{SF}$ versus \Lbol\ for our sample with different symbols for NLS1s and BLS1s. 
We also show the $L_{SF}=\Lbol$ line and the relationship  $L_{SF}=10^{43}\times\left(\Lbol/10^{43}\right)^{0.7} \ergs$
which is adopted from the correlation shown in \cite{Netzer2009}.

\begin{figure*}
\begin{center}$
\begin{array}{cc}
\includegraphics[width=9cm, height=9cm]{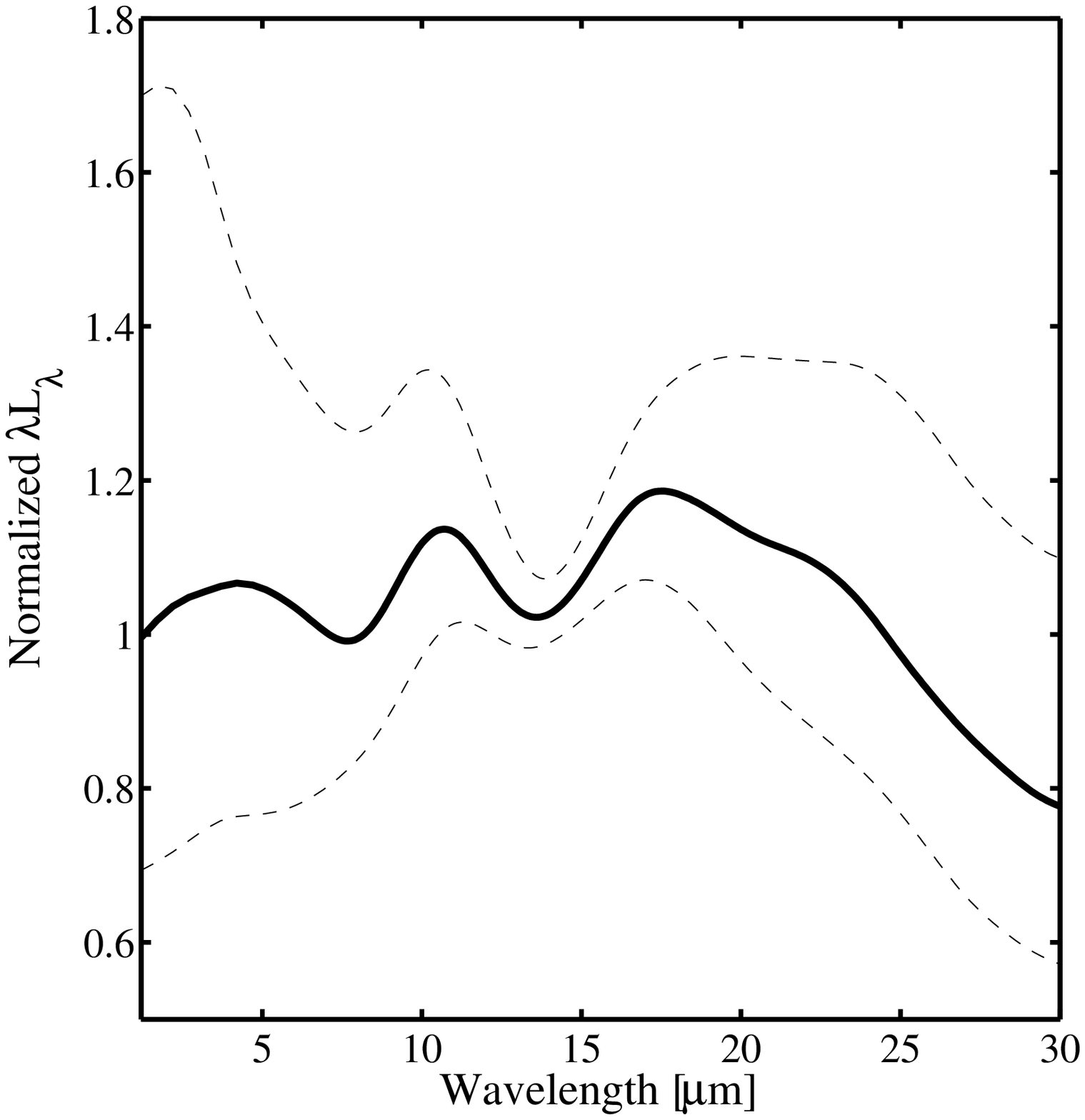} &
\includegraphics[width=9cm, height=9cm]{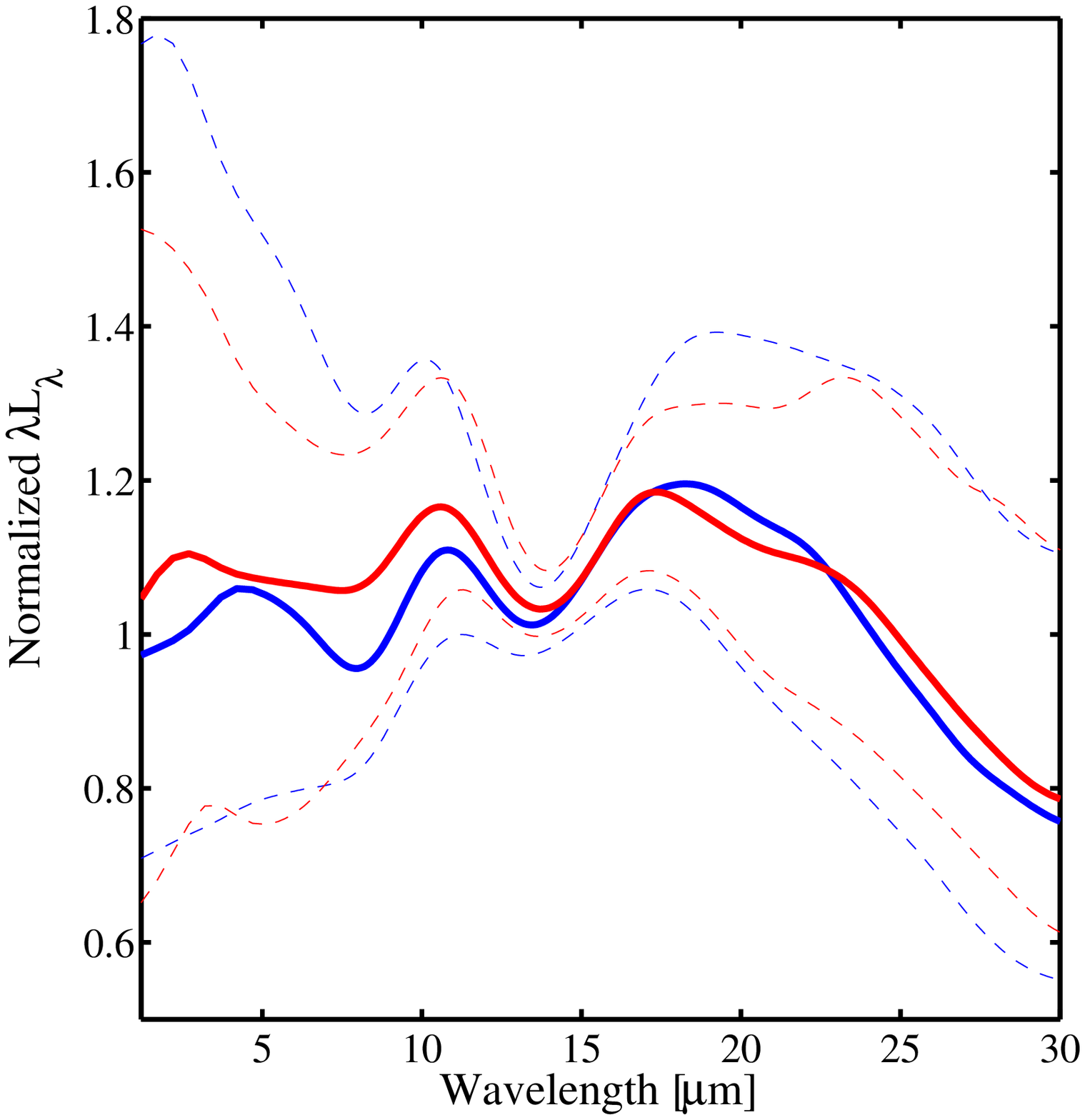}
\end{array}$
\end{center}
\caption{Intrinsic AGN SEDs. Right: median intrinsic AGN SEDs for two sub-groups of NLS1s (blue) and BLS1s (red). 
Dashed lines represent the 25th and 75th percentiles of each sub-group. Left: median intrinsic AGN SED for all 85 sources (see data in Table~\ref{tab:sed}).
As in the right panel, dashed lines represent the 25th and 75th percentiles.
All SEDs have very similar shape and exhibit three peaks at short ($\lesssim5$-\mic) and medium ($\sim10$ and $\sim20$-\mic) wavelengths.
The SEDs are normalized at 14 \mic\ and their shape may be somewhat effected by the chosen normalization point.
}
\label{fig:intrinsicAGN}
\end{figure*}

Altogether, 85 sources (37 NLS1s and 48 BLS1s) have high S/N IRS spectra and AGN dominated MIR SEDs.
For each of these, the range of different templates that meet the PAH criterion is relatively narrow, 
and the uncertainty on the shape of the AGN SED is small.
The SF-template subtracted spectra are normalized at 14-\mic\ and then used 
for constructing median intrinsic AGN SEDs for two groups divided by FWHM(\hb), i.e. NLS1s and BLS1s.
The choice of the normalization point may have some effect on the shape of the median SED.
We choose to normalize the spectra at 14-\mic\ since this wavelength is less effected 
by the broad silicate emission features and the large scatter at longer wavelengths.
The right panel of Figure~\ref{fig:intrinsicAGN} shows the median AGN SED for the two populations of 
narrow- and broad-line AGN in our sample.
The left panel shows the median AGN SED for all 85 sources.
Examining Figure~\ref{fig:intrinsicAGN} it is evident that the general shape of the SEDs, 
regardless of line widths, is very similar.
All SEDs exhibit three distinct peaks at short ($\lesssim5$-\mic) and medium ($\sim10$ and $\sim20$-\mic) wavelengths.
The 10 and 20-\mic\ peaks are likely dominated by silicate-dust emission.
The median intrinsic AGN SED for all 85 sources is also listed in Table~\ref{tab:sed}.
We also examined the median AGN SEDs for two luminosity groups
defined by the median luminosity of the sample at 6-\mic\ ($10^{44.2}$ \ergs).
We do not find any significant difference between the two SEDs.

\begin{table}
  \caption{Median Intrinsic AGN SED}
  \label{tab:sed}
  \begin{center}
    \begin{tabular}{cc} \hline \hline
    Rest Wavelength & Intrinsic AGN SED \\
    (\mic) & $\lambda$L$_\lambda$ (arb.units) \\
    \hline
    0.51 &  0.8401 \\
    1.2  &  0.9956 \\
    1.7  &  1.0182 \\
    2.2  &  1.0364 \\
    2.7  &  1.0479 \\
    3.2  &  1.0555 \\
\hline
  \multicolumn{2}{l}{Table~\ref{tab:sed} is published in its entirety in the electronic edition}  \\
    \end{tabular}
  \end{center}
\end{table}

Our result is consistent with the shape found by \cite{Netzer+07_QUEST2} for the QUEST sample.
The QUEST sample represents the high end of the AGN luminosity range of our current sample, and mostly
consists of broad line sources. \cite{Netzer+07_QUEST2} also divided the QUEST sample into two groups 
of strong and weak FIR emitters. Again, the shapes of the AGN SEDs were remarkably similar.
\cite{Deo2009} employed a similar subtraction method to a large sample of type 1 and 2 Seyfert galaxies.
These authors find that the AGN continuum emission drops rapidly beyond $\sim20$-\mic\ for all AGN types, regardless of SF activity.
Recently, \cite{Mullaney2011} employed a somewhat different method to determine the intrinsic AGN SED of 
a medium-size sample of type-I and type-II AGN.
These authors fit simultaneously the IR SEDs using a SF template and a broken power law that represents the AGN emission.
They find that the warm dust peak is located at somewhat longer wavelengths, between $\sim15$ and 45-\mic,
before dropping rapidly towards FIR wavelengths. This result is not confirmed by our analysis probably due
to the different way of accounting for the SF contribution, and the different nature of the MIR ingredients they consider (a power-law).
We suspect that it may also be affected by the presence of type-II
AGN in their sample and the difficulty to properly account for the host contribution in these sources.

\subsection{Hot Dust Properties}
\label{subsec_dis_hot_dust}

A major goal of this work is to explain the NIR emission in type-I AGN 
and to identify the physical properties of the component responsible for this emission.
The hot dust models used here provide both the continuum emission, which peaks at NIR wavelengths,
and the UV-NIR lines emitted by this component.

\subsubsection{Continuum emission and distances}
\label{subsubsec_dis_hot_dust_continuum_emission_and_distances}

\begin{figure}
\includegraphics[width=9cm, height=9cm]{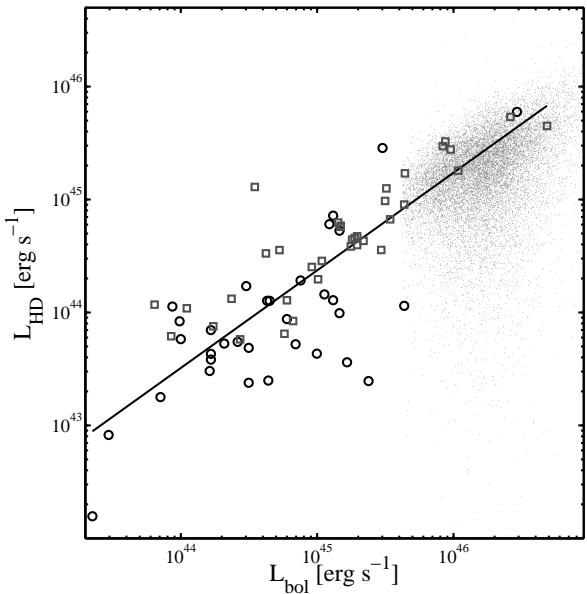}
\caption{Luminosity of the hot dust component (\LHD) vs. AGN bolometric luminosity (\Lbol).
Results from the current work are shown as circles and squares (NLS1s and BLS1s, respectively).
Results for a much larger sample of high luminosity QSOs from \citet{Mor2011} are shown in grey.
The black line represents a simple least squares fit to the data of the current work.
The \woNIR\ sources without NIR data are not included in this plot}
\label{fig:LHD_Lbol}
\end{figure}

The most important parameter of the hot dust component determined by the fitting procedure is its luminosity (\LHD). 
Since the hot dust is optically thick at all UV-optical wavelengths, \LHD\ is simply a measure 
of the covering factor of this component, \CFHD.
Figure~\ref{fig:LHD_Lbol} shows a strong correlation between \LHD\ and \Lbol.
NLS1s (blue circles) exhibit a similar trend to that of BLS1s (red circles) although the scatter in this group is larger.
The solid line in Fig.~\ref{fig:LHD_Lbol} represents a simple least squares fit to the data of the current work with a slope of 0.9.
Fig.~\ref{fig:LHD_Lbol} also shows the results of \cite{Mor2011} (grey dots). These authors fitted the 
(observed) optical to MIR SEDs of $\sim15000$ high luminosity QSOs and measured the luminosity of the hot dust component. 
The relation between \LHD\ and \Lbol, over a large luminosity range, indicates that \CFHD\
in type-I AGN spans a relatively limited range ($\sim$0.1--0.2).
The NLS1s that lie much below the relation of Fig.~\ref{fig:LHD_Lbol} may represent sources 
with exceptionally small \CFHD\ (see \S\ref{subsec_dis_covering_factors})
 
\begin{figure}
\includegraphics[width=9cm, height=9cm]{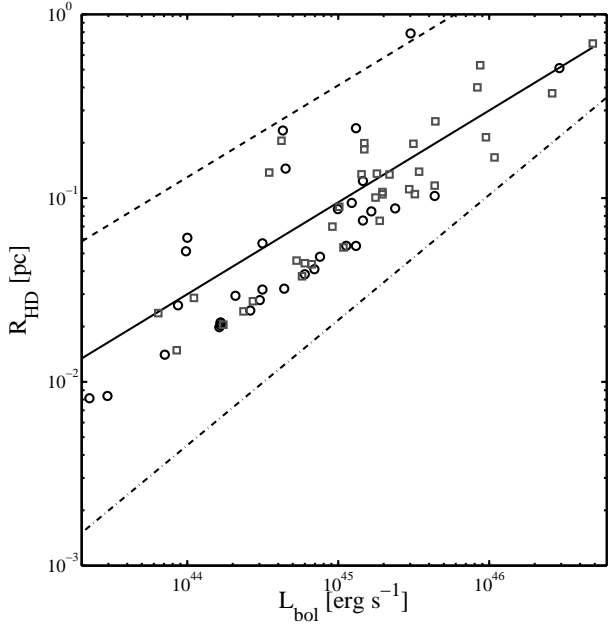}
\caption{Distance to the hot pure-graphite dust components (\RHD) vs. AGN bolometric luminosity (\Lbol)
for two sub-groups of NLS1s (circles) and BLS1s (squares). Dashed and solid black lines represent
the sublimation radii of silicate and pure-graphite dust (eqn. \ref{eq:R_d_Si} and \ref{eq:R_d_c}), respectively.
Dot-dashed line represent the \RBLR-\Lbol\ relation (see text).}
\label{fig:RHotDust}
\end{figure}

The second parameter that is determined by the fitting procedure is the distance from the centre to the hot dust component (\RHD).
Figure~\ref{fig:RHotDust} shows this distance against \Lbol.
The silicate and pure-graphite dust sublimation radii defined by equations \ref{eq:R_d_Si} and \ref{eq:R_d_c}, respectively,
are also shown. We also show the relation between the emissivity-weighted BLR radius for the \Hbeta\ line, \RBLR\ and \Lbol.
This radius is based on an up-to-date version of the \cite{Kaspi2005} relation
taking into account the modifications of \cite{Bentz2009}: $\RBLR = 0.35 \times \left(\Lop/10^{46}\,\ergs \right)^{0.62}\,pc$. 
Fig.~\ref{fig:RHotDust} demonstrates that the clouds are clearly situated outside the dust-free BLR and 
inside the edge of the ``standard'' silicate-dust torus.

Several other model parameters affect the derived distance, for example the column density of the graphite dust 
(which is related to the gas metallicity) that affect the mean dust temperature in an optically thick clouds.
However, changing the values of these parameters do not significantly alter the results found here.
For example, assuming solar metallicity, instead of 2 \Zsun, enlarges the derived distances,
for all sources, by about a factor of 1.25. These distances are still consistent with the assumed location of the dusty clouds.

\subsubsection{The emission line spectrum of dusty BLR clouds}
\label{subsubsec_dis_hot_dust_emission_line}
\begin{figure}
\includegraphics[width=9cm, height=9cm]{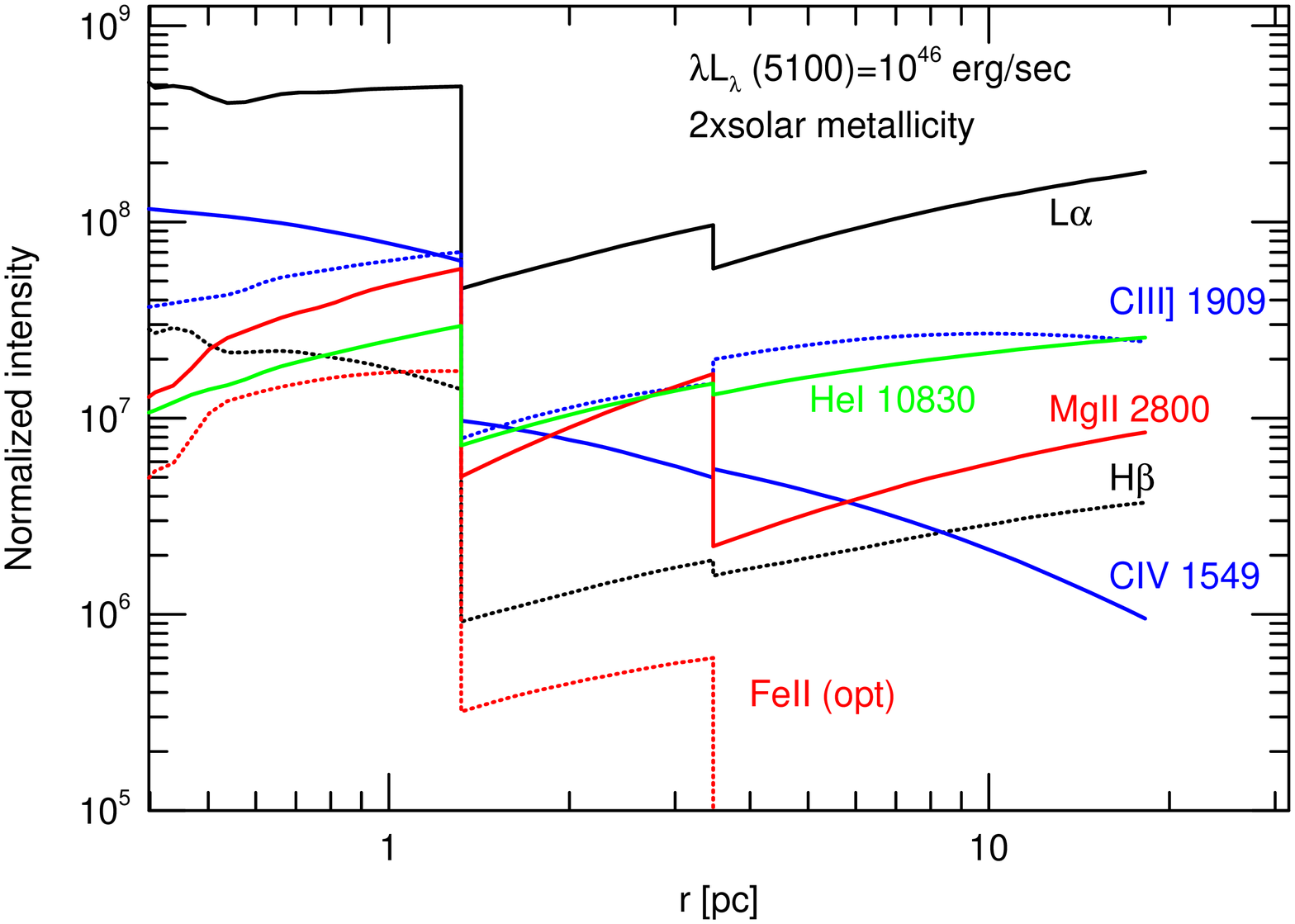}
\caption{Normalized line emissivity per unit covering fraction for strong broad emission
lines for the case of twice solar metallicity. The graphite sublimation radius in this case is at 1.32 pc 
and the silicate sublimation radius at 3.47 pc. These distances are clearly visible due to the large
drop in the emissivity of all lines resulting from the absorption of the ionizing
radiation by the grains and the selective depletion of most metals.}
\label{fig:dust_line_emissivity_2solar}
\end{figure}

So far we have only discussed the hot dust continuum emission.
However, these ``dusty-graphite-BLR'' clouds also produce broad emission lines that will be observed in
the UV-NIR part of the spectrum. This aspect has never been investigated and
is likely to result in more observational constraints on the hot dust model.
Fig.~\ref{fig:dust_line_emissivity_2solar} shows normalized line fluxes, per unit covering factor,
for the 2 \Zsun\ metallicity BLR model considered here. The line intensities are shown as a function of distance for 
the specific case of $\Lop=10^{46} \ergs$. Other cases can be obtained from the diagram by scaling $r$ with $L^{1/2}$

\begin{figure}
\includegraphics[width=9cm, height=9cm]{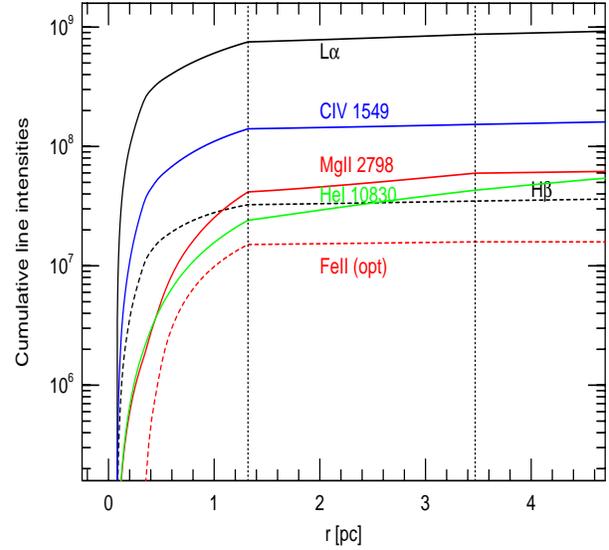}
\caption{Cumulative emission line intensities as a function of distance from the central source 
for the model presented in Fig.~\ref{fig:dust_line_emissivity_2solar}.
The dashed vertical lines represent the graphite and silicate dust sublimation radii for the model.
For most lines, the increase in total emitted flux is very small although not completely zero (see text for details).
}
\label{fig:cumulative_2solar}
\end{figure}

Given the known radial changes in the differential covering fraction, $dC(r)$, we can integrate over the line emissivity to obtain the
cumulative line luminosities produced by the dusty-BLR clouds. This is shown in Fig.~\ref{fig:cumulative_2solar}.
The diagram illustrates the large increase in line flux up to the graphite sublimation radius and the much slower increase beyond this boundary.
For the model in question, the covering fraction of the dusty-graphite-BLR clouds is almost exactly
the same as the cumulative covering fraction of the dust-free clouds. 
If the dusty-graphite-BLR clouds were dust free, the total emitted flux of lines such as \Lya\ and \hb, would roughly double
over this region.
Since much of the ionizing radiation is absorbed by the graphite grains, the increase in intensity is significantly reduced. 
For example, in the dusty-graphite-BLR model considered here, the increase in \Lya\ is 11\%, in \hb\ 7\%, in \CIV\ 6\%, in \MgII\ 38\%,
in \HeINIR\ 69\%, and in \feii\ only 3\%.
Thus, the only significant contributions to the total line intensities are due to \mgii\ and \hei\ lines.
 
The BLR model presented here is not entirely consistent with the SED fitting procedure of \S\ref{sec_spec_decomp}
since all clouds in the pure-graphite zone are contributing to the emission lines yet the SED fitting takes into account
only one such cloud. As illustrated in Fig.~\ref{fig:hot_dust_grid}, this is not a severe limitation since the 
covering-factor-weighted \textit{mean} spectrum has an SED that is very similar to that produced by a single cloud.

The BLR models considered here do not represent the entire range of possible properties.
In particular, the effect of the internal dust on the line emission depends on the ionization parameter ($U$) since
at lower $U$, the fraction of the ionizing continuum absorbed by the dust is reduced. 
This can result in a larger increase of the line intensity in this zone. 
The exact values of $U$, the gas density, and column density of the cloud model are determined by the 
parameter $s$ (\S\ref{subsec_spec_decomp_hot_dust}).
A more detailed consideration of these possibilities is deferred to a future publication.

\subsection{Torus Properties}
\label{subsec_dis_torus}

As explained in \S\ref{subsec_spec_decomp_torus}, the presence of hot dust clouds 
at the inner edge of the torus will affect the torus emission spectrum.
The hot dust clouds will absorb a large fraction of the incident UV radiation and re-emit it at NIR wavelengths
and at different directions than the original AGN radiation. This will change the SED and total luminosity incident on the torus.

As a first attempt to solve this issue without resorting to a full new torus model (which is not yet available),
we subtracted the hot-dust luminosity, \LHD, from \Lbol\ and used this value in order to calculate the torus models
(see \S\ref{subsec_spec_decomp_torus}).
For this reduced amount of incident luminosity, we used the exact same scheme as M09 in which the
torus normalization is not a free parameter but rather set by the (now reduced) AGN luminosity.
We found that most sources still could not be fitted using low inclination angle, those representing type-I AGN tori.
To fit the spectra using such low inclination models one would need to \textit{artificially} 
reduce \Lbol\ by a factor that is larger than the one implied by the hot dust emission.
This unsuccessful attempt may imply that the viewing angle of the hot dust (which is not considered here) can play an 
important role or that the geometry of the hot dust clouds is different from that of the torus.
For example, if the hot dust component has similar geometry to that of the torus, 
its emission would also be highly anisotropic. Moreover, if the two components have different inclination
angles, the amount of radiation that the torus would receive from the hot dust component would be 
different than the amount inferred by the observed NIR flux.
We conclude that the only way to fully examine this possibility is to include a pure-graphite dust
component into the radiative transfer calculation of the clumpy torus.

In further attempt to solve this issue, we set the normalization of the torus component to be 
a free parameter of the fitting procedure (\S\ref{subsec_spec_decomp_torus}).
This way, models with low energy output, small physical size, or high inclination angles w.r.t. the line of sight,
fit well the observed spectra. 
This adds a considerable uncertainty to the derived values of the geometrical parameters of the fit.
However, the average sample properties are less likely to be affected.

The following conclusions refer to the distribution of the different torus 
parameters for all sources with good IRS S/N ratio (49 NLS1s and 51 BLS1s).
We found that some parameters have a narrow distribution around a mean value and in
some cases only a single acceptable value. Other parameters exhibit a broader, more uniform distribution.
The parameter distributions of all sources were combined by giving each value within the
acceptable range in a certain source its relative weight in the distribution.
The results are (see M09 for more explanation of the parameters):
\begin{enumerate}
\item
The median torus width parameter is $\sigma=35$.
\item
The radial extent of the torus, Y, has a broad distribution with a median value of 30.
The range in Y implies torus outer sizes that range between $\sim$1.7 to 85 pc.
\item
The average number of clouds along an equatorial ray, $N_{0}$,
has a broad distribution with a median of 4 clouds.
\item
The parameter that specifies the radial power-law distribution of the clouds, q,
has a median value of 1.7. This parameter is related to the anisotropy of the torus radiation.
As q decreases the torus radiation becomes less isotropic.
\item
The median optical depth of a cloud is \tv=38. This parameter has a broad distribution
with \tv$\geq20$. Since the requirement for large MIR optical depth is
\tv$\sim 10$, it is not surprising that the torus models are not very sensitive to the exact value of \tv\ beyond this value.
\item
The torus inclination angle shows a broad distribution for all $i \leq 60^{\circ}$ with a median of 50 $^{\circ}$. 
This again is consistent with the assumption that the direct line-of-sight
to the centre of type-I AGN is almost completely free of obscuring material.
\end{enumerate}
We stress again that we do not attach great significance to a specific value of a specific torus parameter
but rather make sure that all parameters lie within a reasonable range for type-I AGN.

\subsection{NLR Properties}
\label{subsec_dis_NLR}

\begin{figure}
\includegraphics[width=9cm, height=9cm]{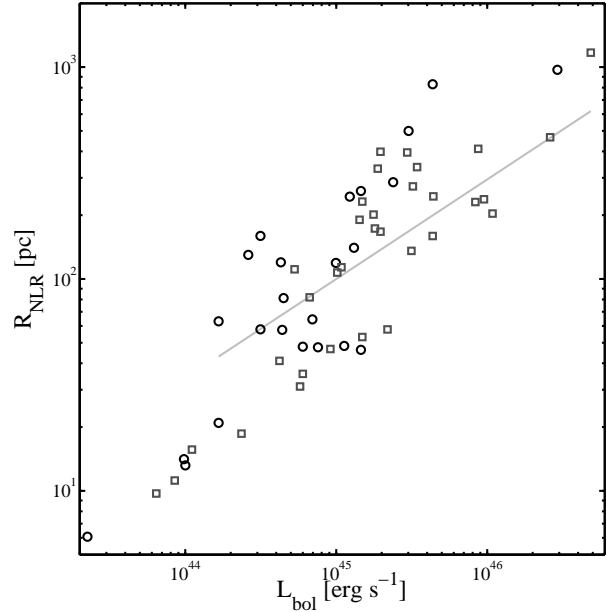}
\caption{NLR cloud distances vs. AGN bolometric luminosity. 
NLS1s are marked by circles while BLS1 are marked by squares.
The slope from M09 (0.47) is shown by the grey line for comparison.
The \lowSN\ sources with very low IRS S/N are not included in this plot.}
\label{fig:RNLR}
\end{figure}

The important parameters of the NLR component are the distance from the centre,
which determines the dust temperature, and the total dust luminosity which determines the fraction of
MIR flux emitted by such clouds. The properties of this component are similar to those described in M09.
Figure~\ref{fig:RNLR} shows NLR cloud distances against the bolometric luminosity of the objects.
Using simple least squares regression we find the following scaling relationship,
\begin{equation}
\RNLR = 460 \times \left( \frac{\Lbol}{10^{46}\, \ergs}\right)^{0.67 \pm 0.05} \,\,  \rm{pc}.
\label{eq:R-NLR_Lbol_excatfit}
\end{equation}
The grey line in Fig.~\ref{fig:RNLR} represents the scaling relation (slope 0.47) found for the QUEST sample by M09.
Although we find a slightly steeper slope, both results are consistent within the observed scatter.
The distances found here are similar to the values found by M09, however the scatter in the current sample is larger.

As explained, the main difference from M09 regarding the NLR component is the more detailed treatment of the SF contribution.
For example, the emission from the dusty NLR component peaks at MIR wavelengths (above $\sim$20-\mic)
where the SF contribution to the observed spectrum is substantial.
M09 used a single M82 template to subtract the observed spectrum. 
For sources with strong SF activity (i.e., those with high \Lbol\ or NLS1s that exhibit strong SF, see S10)
this template does not properly account for the observed FIR emission.
Other SF templates have larger contribution towards shorter wavelengths ($\sim$30-\mic).
Consequently, the NLR component would have less weight in the AGN MIR spectrum and different spectral shape.
This can result in larger NLR distances and smaller luminosity. 
We consider this as an additional uncertainty on the determination of the NLR distance
and its relative contribution.

\subsection{Covering Factors of the Dusty Components}
\label{subsec_dis_covering_factors}

\begin{figure}
\includegraphics[width=9cm, height=9cm]{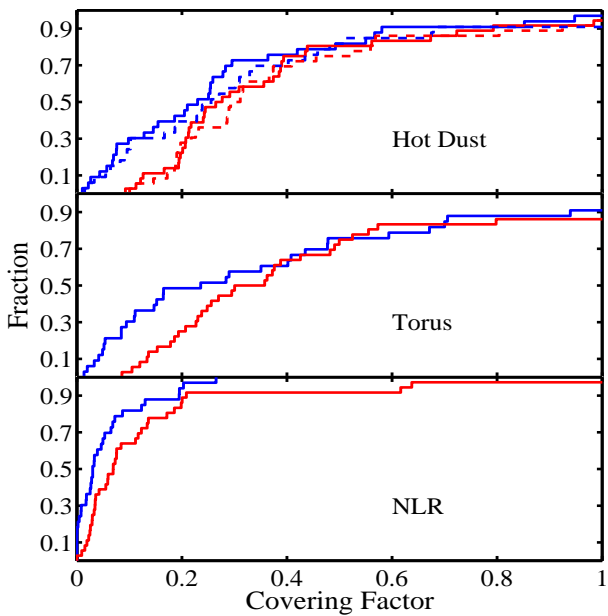}
\caption{Cumulative distribution functions of the covering factors of the three AGN dusty components.
For all components, the covering factors tend to be smaller in NLS1s (blue) than in BLS1(red).
The distribution of \CFHD\ and the torus covering factor are very similar indicating that about
half of the obscuration is due to hot pure-graphite dust clouds which are typically not included in
``standard'' torus models. Dashed lines in the upper panel represent \CFHD\ obtained by fitting the
1--5 \mic\ range using the hot dust component only (without a torus component - see text)}
\label{fig:CF_cdf}
\end{figure}

A major assumption of this work is that the entire MIR spectrum, after starburst subtraction, is reprocessed AGN radiation.
This can be used to deduce the covering factor of the central source by the three components (see also M09 and references therein).
The covering factors are defined by the ratio between the total luminosity of each component and \Lbol.
Figure~\ref{fig:CF_cdf} shows the cumulative distribution functions of the covering factors of the different components.
The blue line represents the NLS1s in the sample and the red line represents the BLS1s.
As seen from the diagram, NLS1s tend to have smaller covering factors compared with BLS1s.
The median value of \CFHD\ for the NLS1s is 0.23, for the BLS1s 0.27, and for the entire sample is 0.25.
The median value of the torus covering factor is 0.24 for the NLS1s and 0.33 for the BLS1s.
The covering factor of the NLR component has a median value of 0.03 for the NLS1s and 0.07 for the BLS1s.

S10 found that NLS1s tend to have stronger SF activity compared with BLS1s at a given AGN luminosity.
Since we fit the SF subtracted SED this may effect the derived torus and NLR covering factors.
However, this cannot explain the difference in the covering factors of the hot dust component since the 
SF contribution to the 1--5 \mic\ wavelength range is very small. 
Host contribution to the optical range (i.e. to \Lop) would result in an overestimation of the AGN bolometric luminosity and 
consequently lower the estimated covering factor (see discussion on the uncertainty of \Lbol\ in \S\ref{sec_sample}).
Such host contribution to the optical spectrum however, is more significant in lower luminosity AGN regardless of their line widths,
and cannot explain the observed discrepancy. 

We further checked the influence of the torus component on the fitting of the NIR wavelength range (1--5 \mic)
where the hot dust component dominates the spectrum. We fitted this limited wavelength range using 
a hot-dust component only, without including a torus component. 
The dashed lines in the upper panel of Fig.~\ref{fig:CF_cdf} represent the \CFHD\ obtained in this manner.
The median value of this \CFHD\ for all sources is 0.29 and is slightly higher than that found by the 
fitting of the entire SED using all three components (0.25).
Thus using NIR photometry \textit{only}, \CFHD\ (and \LHD) are overestimated, on average, by 16\%.

\begin{figure*}
\includegraphics[width=1\textwidth]{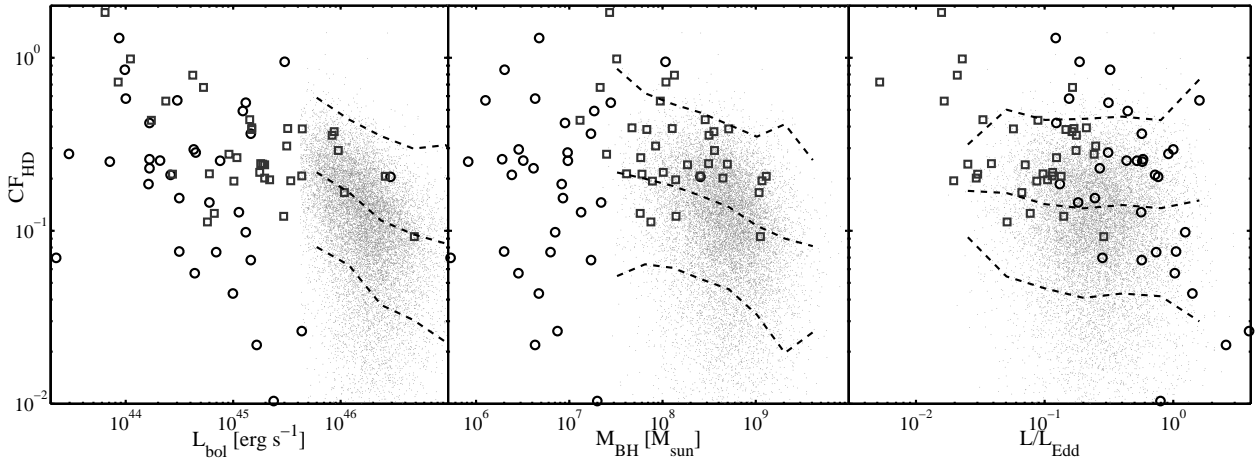}
\caption{Covering factor of the hot pure-graphite dust component against \Lbol\ (left), \mbh\ (middle), and \Ledd\ (left panel).
In all panels, NLS1s are represented by the circles and BLS1s by squares. 
Results of \citet{Mor2011} (grey dots) are shown for comparison. 
Dashed black lines represent the peak of the \CFHD\ distribution and the 99th percentiles boundaries - see text}
\label{fig:CFHD_Lbol_MBH_Ledd}
\end{figure*}

In order to explore the relation with AGN properties we focus on \CFHD\ and compare it against \Lbol, \mbh, and \Ledd.
Fig.~\ref{fig:CFHD_Lbol_MBH_Ledd} shows \CFHD\ against these properties for the current sample of NLS1s (blue circles)
and BLS1s (red circles). The results of \citet[grey dots]{Mor2011} are shown for comparison.
The black dashed lines in Fig.~\ref{fig:CFHD_Lbol_MBH_Ledd} represent the 99th percentiles boundaries of the
\CFHD\ distribution of the \cite{Mor2011} sample.
These percentiles were calculated by assuming that the \CFHD\ distribution, in a certain luminosity bin, should be symmetrical around
the peak value and by mirroring the high-\CFHD\ side of the distribution. This is done in each 0.2 dex luminosity bin separately.
\cite{Mor2011} suggest that all the points which lie below the lower dashed lines can be regarded as hot-dust-poor AGN.
The fraction of such sources in their sample is $\sim$20\% and does not depend on source luminosity.

The left panel of Fig.~\ref{fig:CFHD_Lbol_MBH_Ledd} presents a clear anti-correlation, 
in the sense of a decreasing \CFHD\ with increasing \Lbol. 
This trend is further confirmed by both Pearson's and Spearman's rank correlation tests (p value $\ll0.01$ for both tests).
Several earlier studies suggested a similar trend \cite[e.g.,][]{Wang2005, Maiolino2007a, Treister2008},
however these were based on the total MIR emission which translates to the covering factor of the entire dusty 
structure, and not just the hot dust.
\cite{Gallagher2007} suggest that this \CFHD-\Lbol\ anti-correlation may be the manifestation of dust extinction in the host galaxy
\footnote{\cite{Gallagher2007} present a somewhat different covering factor since they integrate over a larger range 
of IR wavelengths ($\sim$1--100 \mic).}
\cite{Mor2011}, however, showed that such an anti-correlation is still very significant even for the bluest sources in their sample.
These ``blue'' sources have optical slopes greater than $-0.4$, and are less likely to be affected by extinction.
The physical mechanism responsible for the decrease of covering factor with \Lbol\ is still undetermined.
One possibility is a ``receding torus'' scenario \citep{Lawrence1991}, where higher luminosity implies larger
dust sublimation distance, and hence an obscuring structure that is located farther away from the centre.
In this scenario, the geometry of the hot-dust clouds must be toroidal and have a constant height.
Another possibility is that the \CFHD-\Lbol\ anti-correlation is analogous to the anti-correlation found 
between the equivalent width of different BLR lines (\Hbeta\ and \civ) and AGN luminosity \cite[e.g.,][]{Baldwin1977, Netzer2004, Baskin2005}.
The \CFHD-\Lbol\ anti-correlation cannot explain the lower \CFHD\ of the NLS1s in the current sample, 
since both BLS1s and NLS1s sub-groups span a similar luminosity range.

The NLS1s in our sample have lower \mbh\ than the BLS1s.
As can be seen in the middle panel of Fig.~\ref{fig:CFHD_Lbol_MBH_Ledd}, most of 
the sources with $\log \left (\mbh/\Msun \right ) < 7.3$ are NLS1s and $\sim$20\% of 
them have \CFHD\ lower than 0.1. 
No significant correlation was found between \CFHD\ and \mbh.
If the NLS1s that have very low \CFHD\ are truly hot dust poor AGN it may suggest that these
sources are more common in low \mbh\ sources.
\cite{Mor2011} did not find such a dependence of the fraction of hot dust poor AGN on \mbh.
However, the reason that this dependence was not found can be related to the relatively narrow \mbh\ range
that the \cite{Mor2011} sample span.

\Ledd\ is known to be higher in NLS1s. The right panel of Fig.~\ref{fig:CFHD_Lbol_MBH_Ledd} suggests an
anti-correlation between \CFHD\ and \Ledd, over two orders of magnitude in \Ledd.
The NLS1s that lie below the \CFHD-\Lbol\ and \CFHD-\mbh\ relations are also the sources with the highest \Ledd. 
\cite{Mor2011} did not find significant correlation between \CFHD\ and \Ledd. However, this is probably due to the relatively
narrow range of \Ledd\ in their sample.
Similar to the relation with the AGN luminosity, \cite{Netzer2004} and \cite{Baskin2005} found that the equivalent width of 
the broad \Hbeta\ and \civ\ lines also anti-correlates with \Ledd.

\section{Conclusions}
We conducted a detailed investigation of the NIR-FIR SED of a large sample of Spitzer-observed AGN.
We fitted the spectra of 51 NLS1s and 64 BLS1s using a three component model: a clumpy torus, a dusty NLR, and 
a hot pure graphite-dust component which is a continuation of the BLR.
The fitting was performed on SF subtracted SEDs using SF templates that take into 
account the entire range of possible host galaxy properties. 

The main results of the investigation are:
\begin{itemize}
\item 
The intrinsic MIR AGN SED is very similar in NLS1s and BLS1s regardless of AGN luminosity.
All SEDs exhibit three distinct peaks at short ($\lesssim5$-\mic) and medium ($\sim10$ and $\sim20$-\mic) wavelengths.
The first peak is likely dominated by emission from pure-graphite dust located outside the dust-free BLR and 
within the ``standard'' silicate-dust torus.
The other two peaks are likely dominated by silicate-dust, corresponding to the known 10 and 18-\mic\ broad emission features.
The long wavelength downturn of the AGN SED is at about 20--25 \mic.

\item
Our detailed modelling of the hot-dust component allows us to explore the emission line spectrum of such clouds.
Most line emission is dramatically suppressed in these dusty clouds.
However, the \MgII\ and \hei\ lines are still contributing significantly to the total BLR spectrum. 

\item
We calculated the covering factors of all AGN-related component and found that the covering factor of the 
hot-dust component is about 0.25. The covering factor of the torus component is comparable 
with a median value of 0.28. The NLR has a much smaller covering factor of about 0.07.
The NLS1s in our sample tend to have smaller covering factors than BLS1s.

\item
The covering factor of the hot-dust component, \CFHD, anti-correlates with both \Lbol\ and \Ledd.
The \CFHD-\Lbol\ anti-correlation may be related to the idea of a ``receding torus'' where higher luminosity implies larger
dust sublimation distance, and hence an obscuring structure that is located farther away from the centre.
This however requires that the hot dust component would have a toroidal structure with constant height.
The origin of the \CFHD-\Ledd\ relation is unknown and is not a consequence of the \CFHD-\Lbol\ relation
since the NLS1 and BLS1 sub-samples have similar distribution of \Lbol.
Both the \CFHD-\Lbol\ and the \CFHD-\Ledd\ relations may be analogues of the anti-correlations found between 
the equivalent widths of broad \Hbeta\ and \civ\ lines and AGN luminosity and \Ledd.

\end{itemize}


\section*{Acknowledgments}

We are grateful to Benny Trakhtenbrot and Ido Finkelman for useful discussions.
We thank the DFG for support via German-Israeli Cooperation grant STE1869/1-1.GE625/15-1.
Funding for this work has also been provided by the Israel Science Foundation grant 364/07.
This publication makes use of data products from the Two Micron All Sky Survey, which is a joint project of the 
University of Massachusetts and the Infrared Processing and Analysis Center/California Institute of Technology, 
funded by the National Aeronautics and Space Administration and the National Science Foundation.


\label{lastpage}
\end{document}